\documentclass[10pt]{article}

\usepackage{natbib}
\usepackage{textcomp}
\usepackage{apalike}
\usepackage{amsmath}
\usepackage{hyperref}
\usepackage{graphicx}
\usepackage{array}
\usepackage{booktabs}

\begin{document}

\title{RhoanaNet Pipeline: Dense Automatic Neural Annotation}

\author{
Seymour Knowles-Barley
\thanks{Department of Molecular and Cellular Biology and Center for Brain Science,\newline Harvard University.}
\thanks{Present Address: Google Inc.}
\,
Verena Kaynig
\thanks{School of Engineering and Applied Sciences, Harvard University.}
\,
Thouis Ray Jones
\footnotemark[1]
\footnotemark[3]
\\
Alyssa Wilson
\footnotemark[1]
\,
Joshua Morgan
\footnotemark[1]
\thanks{Present Address: Washington University in St. Louis}
\,
Dongil Lee
\footnotemark[1]
\,
Daniel Berger
\footnotemark[1]
\\
Narayanan Kasthuri
\footnotemark[1]
\thanks{Present Address: Argonne National Laboratory.}
\,
Jeff W. Lichtman
\footnotemark[1]
\,
Hanspeter Pfister
\footnotemark[3]
}

\maketitle

\begin{abstract}
Reconstructing a synaptic wiring diagram, or connectome, from electron microscopy (EM) images of brain tissue currently requires many hours of manual annotation or proofreading \citep{Kasthuri2010Neurocartography, Lichtman2008Ome, Seung2009Reading}. The desire to reconstruct ever larger and more complex networks has pushed the collection of ever larger EM datasets. A cubic millimeter of raw imaging data would take up 1 PB of storage and present an annotation project that would be impractical without relying heavily on automatic segmentation methods.
The RhoanaNet image processing pipeline was developed to automatically segment large volumes of EM data and ease the burden of manual proofreading and annotation. Based on \citep{Kaynig2015Largescale}, we updated every stage of the software pipeline to provide better throughput performance and higher quality segmentation results. We used state of the art deep learning techniques to generate improved membrane probability maps, and Gala \citep{NunezIglesias2014Graphbased} was used to agglomerate 2D segments into 3D objects.

We applied the RhoanaNet pipeline to four densely annotated EM datasets, two from mouse cortex, one from cerebellum and one from mouse lateral geniculate nucleus (LGN). All training and test data is made available for benchmark comparisons. The best segmentation results obtained gave $V^\text{Info}_\text{F-score}$ scores of 0.9054 and 09182 for the cortex datasets, 0.9438 for LGN, and 0.9150 for Cerebellum.

The RhoanaNet pipeline is open source software. All source code, training data, test data, and annotations for all four benchmark datasets are available at \href{http://www.rhoana.org}{www.rhoana.org}
\end{abstract}

\section{Introduction}
Reconstructing the wiring diagram of the nervous system at the level of single cell interactions is necessary for discovering the underlying architecture of the brain and investigating the physical underpinning of cognition, intelligence, and consciousness \citep{Kasthuri2010Neurocartography, Lichtman2008Ome, Seung2009Reading}. Advances in the data acquisition process now make it possible to image millions of cubic microns of tissue with hundreds of TB of raw data \citep{Kasthuri2015Saturated}. With these techniques, a cubic millimeter of raw imaging data would take up 1 PB of storage and present an annotation project that would be impractical without relying heavily on automatic
segmentation methods \citep{Kaynig2015Largescale}.
Currently the general development of new image segmentation and reconstruction techniques for Connectomics is hindered by the lack of available benchmark data sets. These benchmarks are standard practice in computer vision and machine learning research and driving the development of new methods \citep{Krizhevsky2009CIFAR, Martin01BerkleyData}. The benefit of benchmarks for Connectomics became evident when the ISBI 2012 competition with one publicly released data set greatly improved the state of the art by opening the field to broader range of computer vision research \citep{ArgandaCarreras2015ISBI}.

\begin{figure*}[!tpb]
\centerline{\includegraphics[width=\textwidth]{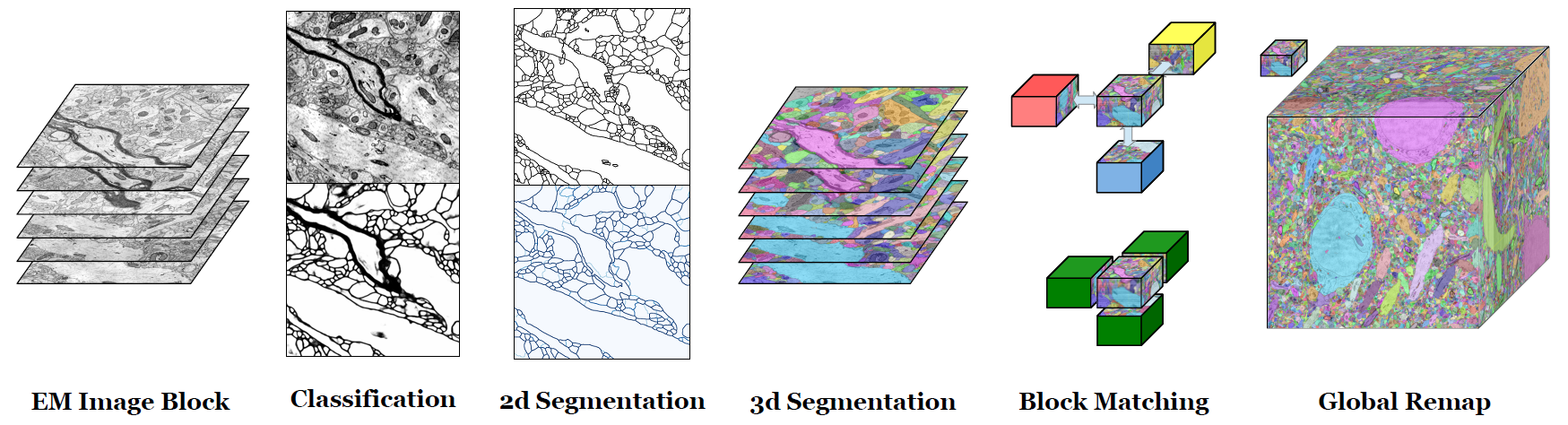}}
\caption{The RhoanaNet segmentation pipeline. First we generate boundary probability maps for each electron microscopy image. These are then used to obtain a 2D segmentation of neuronal regions, followed by region agglomeration across sections to obtain a 3D segmentation. All steps are embarrassingly parallel and can be executed in blocks of image volumes. The 3D segmentations from different blocks are then merged into one global reconstruction volume. }\label{fig:pipeline}
\end{figure*}

\begin{figure*}[!tpb]
\centerline{\includegraphics[width=\textwidth]{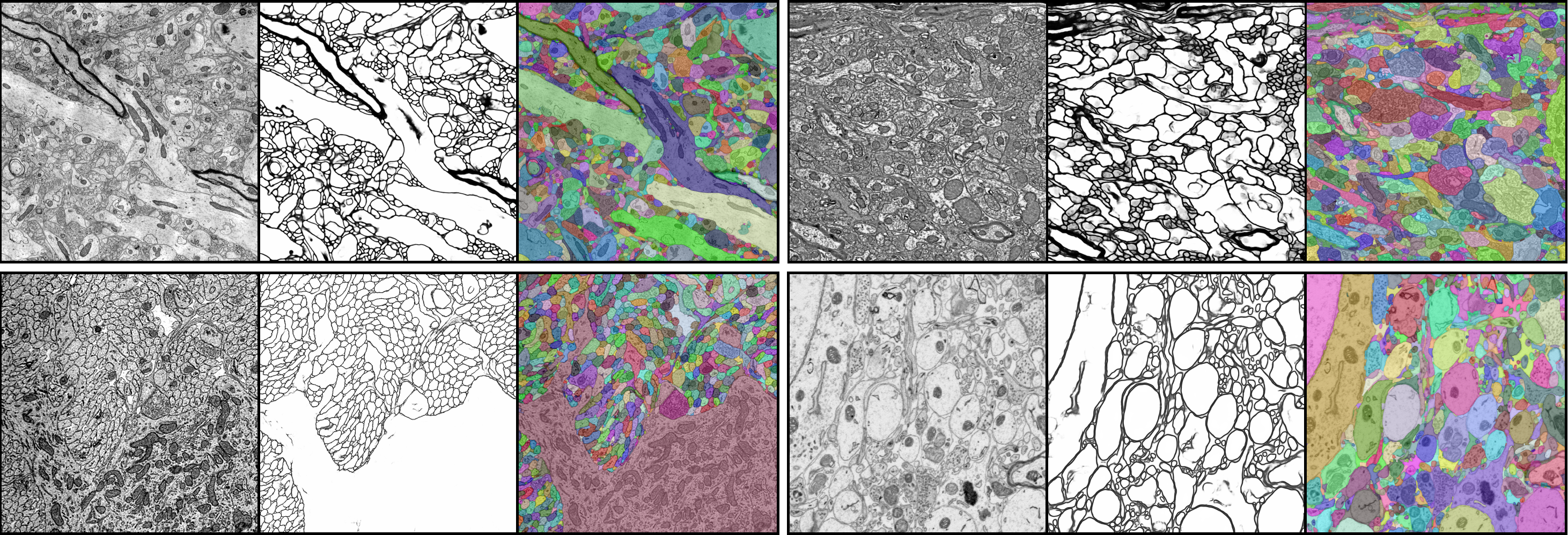}}
\caption{Image summary of the four datasets. Clockwise from top left with image heights: S1 Cortex (12.29 microns), LGN (16.38 microns), Cerebellum (8.19 microns), ECS Cortex (8.19 microns). Each panel consists of a source EM image (left), U-Net membrane probability results (middle) and final segmentation results (right). All images are from test volume data.}\label{fig:case_studies}
\end{figure*}

In this paper we present a complete framework for benchmarking Connectomics reconstructions and apply it to four data sets with annotations (see figure \ref{fig:case_studies}), together with an improved version of our automatic segmentation pipeline called RhoanaNet (see figure \ref{fig:pipeline}), evaluation metrics and our current benchmark results. 

The RhoanaNet pipeline and segmentation proofreading tools Mojo and Dojo \citep{Haehn2014Design} are open source software. Source code and data are available online at \href{http://www.rhoana.org}{www.rhoana.org}.

\section{Methods}
We first describe the four benchmark data sets and then our improved automated pipeline and the evaluation metrics.

\subsection*{RhoanaNet Pipeline}
Since the original publication of the Rhoana pipeline \citep{Kaynig2015Largescale}, we have updated every stage to take advantage of state-of-the-art deep learning techniques, resulting in large improvements in quality and performance. The new RhoanaNet Pipeline consists of five main stages. An overview is given in figure \ref{fig:pipeline}. The first stage of the pipeline is membrane classification, where cell membranes are identified in the 2D images producing a membrane probability map. Next, 2D candidate segmentations are generated based on the membrane probability for each image. These segmentations are then grouped across sections into geometrically consistent 3D neuron reconstructions. For this stage, 3D blocks are cropped from the full volume and each block is processed individually to produce a 3D segmentation. An optional cleanup stage removes very small objects and objects completely surrounded by a single object. In the fourth stage, blocks are matched pairwise with neighboring blocks, and overlapping objects are joined. Finally, matched objects are joined globally to produce a consistent segmentation for the full volume.

The modular pipeline approach allows each step to be improved or replaced independently of the rest of the pipeline. This is particularly useful for a direct comparison between methods which only address a part of the pipeline, e.g. updating the membrane classifier or the region agglomeration stage. 

In comparison to the original Rhoana pipeline \citep{Kaynig2015Largescale}, the membrane classification stage has been updated to use state-of-the-art deep learning techniques, the region agglomeration has been changed from segmentation fusion \citep{VazquezReina2011segmentationFusion} to Gala \citep{NunezIglesias2014Graphbased} and the 2D segmentation and pairwise matching stages were updated to be more efficient. In addition, the pipeline was transitioned from MATLAB to a C++ and Python code-base which resulted in run-time performance improvements. The code for our pipeline can be found at \href{http://www.rhoana.org}{www.rhoana.org}.

\subsection*{Deep Learning}
We used deep learning techniques to generate improved membrane probability maps and 2D segmentations. The Keras deep learning library, \citep{chollet2015keras} with the Theano back-end \citep{Theano2016}, was used to train a U-Net network \citep{Ronneberger2015} to predict the probability of a pixel in the electron microscopy image representing a cell boundary membrane. The U-Net architecture follows the defaults described by Ronneberger et al. It consists of layer blocks containing three convolutional layers plus either a max-pooling layer for down-sampling or a deconvolutional layer for up-sampling. The network first has four blocks downsampling the input and then four blocks upsampling again to the original resolution. Residual connections are built between downsampling and upsampling blocks at matching resolutions. Ronneberger et al. showed that this network architecture is very suitable to membrane detection in electron microscopy images, while still maintaining a good throughput rate. On a GTX 970 GPU the data throughput rate is 1 megapixel per second; about two orders of magnitude faster than the random forest classification on a CPU.
The U-Net architecture is sensitive to contrast variations in the images. Therefore, all network training and test data were pre-processed using CLAHE.

\subsection*{2D Segmentation}
With improved classification performance from the deep networks the 2D segmentation stage no longer requires a graph-cut gap completion step \citep{Kaynig2015Largescale}. Instead, a simple watershed on Gaussian smoothed membrane probability maps is used to generate an over-segmentation, which then serves as input for the region agglomeration stage. 

\subsection*{Region Agglomeration}
The over-segmentation obtained from the previous step needs to be grouped into geometrically consistent 3D objects of neuronal structures. Gala \citep{NunezIglesias2014Graphbased} uses a random forest classifier to predict the probability of two segments belonging to the same object. These scores are then used in an agglomerative clustering scheme to group the segmented regions. The random forest region agglomeration classifier is not only trained to group the initial segments, but also for later stages during the agglomeration phase. Iglesias et al. showed that this form of training significantly improves the clustering result. Compared to the previously employed segmentation fusion, Gala performs with greater accuracy for areas with branching structures such as dendritic spines in the cortex. For volumes with fewer branches, segmentation fusion and Gala perform with about the same accuracy.

\subsection*{Pairwise Block Matching}
Pairwise matching was previously solved using Binary Integer Linear Program optimization \citep{VazquezReina2011segmentationFusion}. We have replaced this step with a much faster algorithm based on the stable matching algorithm \citep{Gale1962College}. Block matching is performed using multiple image planes from the overlapping volume between neighboring blocks. Objects are first matched by the stable matching algorithm for objects with overlaps above a given threshold (usually set to approximately $100nm^2$ per overlapping slice). After the first round of stable matching, any remaining unmatched objects are optionally matched to their largest partners.

\subsection*{Datasets}
We provide four benchmark datasets, each of which presents different challenges for segmentation.

\subsubsection*{S1 Cortex} This dataset was collected from mouse somatosensory cortex \citep{Kasthuri2015Saturated} and consists of several ground truth annotated volumes. The data is publicly available on the Open Connectome website (\href{http://www.openconnectomeproject.org}{www.openconnectomeproject.org}) and has been partly analyzed in an open benchmark competition (\href{http://brainiac2.mit.edu/SNEMI3D/home}{brainiac2.mit.edu/SNEMI3D/home}). The data consists of 3 fully annotated sub-volumes from the whole S1 data set described in Kasthuri's paper. This data has a resolution of 6 nm per pixel and a section thickness of about 30 nm. AC4 is a fully annotated volume of size $1024 \times 1024 \times 100$ pixels and corresponds to the training data of the benchmark competition from 2013. In addition we provide a similar sized volume AC3 consisting of $1024 \times 1024 \times 300$ pixels and a large cylindrical volume of a reconstructed dendrite and surrounding structures. The cylindrical volume fits inside a $2048 \times 2048 \times 300$ pixel crop from the S1 dataset. We used half of AC3 and all of AC4 as training data for both membrane classification and agglomeration stages of the pipeline, and the large dendrite volume as test data. All volumes are densely packed with dendrites and axons with many branching structures and small processes such as spine necks which make automatic reconstruction particularly challenging. This image data contains very few artifacts.

\subsubsection*{Cerebellum}
This dataset was collected from developing mouse cerebellum, and contains a mix of parallel fibers and Purkinje cell processes. The parallel fibers generally travel in the same direction without branches, and therefore present a relatively easy task for automatic reconstruction. However, the developing Purkinje cells consist of many branching processes and contain sub-cellular structures that are difficult to differentiate from the outer membrane of cells. 
The resolution of this data is about 8 nm per pixel with a section thickness of 30 nm. 
The data set consists of one fully annotated volume of $1024 \times 1024 \times 100$ pixels. We used the first 50 sections for training and the last 50 sections to evaluate test performance. 

\subsubsection*{LGN}
This training and test volume is a small part of a 67 million cubic micrometer volume from mouse lateral geniculate nucleus (LGN) \citep{Morgan2016Fuzzy}. Cell membranes appear densely packed and contrast is lower than in the other datasets. The data also contains some challenging staining artifacts, which are common in electron microscopy data and need to be addressed by automated reconstruction methods. The resolution of this data set is $4 nm \times 4 nm \times 30 nm$. The training volume is $2360 \times 2116 \times 151$ pixels. The test volume is $2360 \times 2116 \times 20$ pixels.

\subsubsection*{ECS Cortex}
This dataset was collected from mouse cortex, with the tissue processed to preserve the extra-cellular space (ECS) normally present in brain tissue. This data looks very different to conventionally stained EM data, as neuronal regions are not densely packed. It is the smallest data set in our collection, but the unique staining and resulting tissue properties make it a very interesting data set to analyze. 
The resolution of this data is 4 nm per pixel with a section thickness of 30 nm. The training volume is $1536 \times 1536 \times 98$ pixels. A small subsection of this volume consisting of $632 \times 560 \times 40$ pixels has been annotated again by a second person. To train our membrane detection network we used the training images and annotations from both label stacks. The test volume consists of $1536 \times 1536 \times 20$ pixels.  

\begin{figure*}[!tpb]
\centerline{\includegraphics[width=\textwidth]{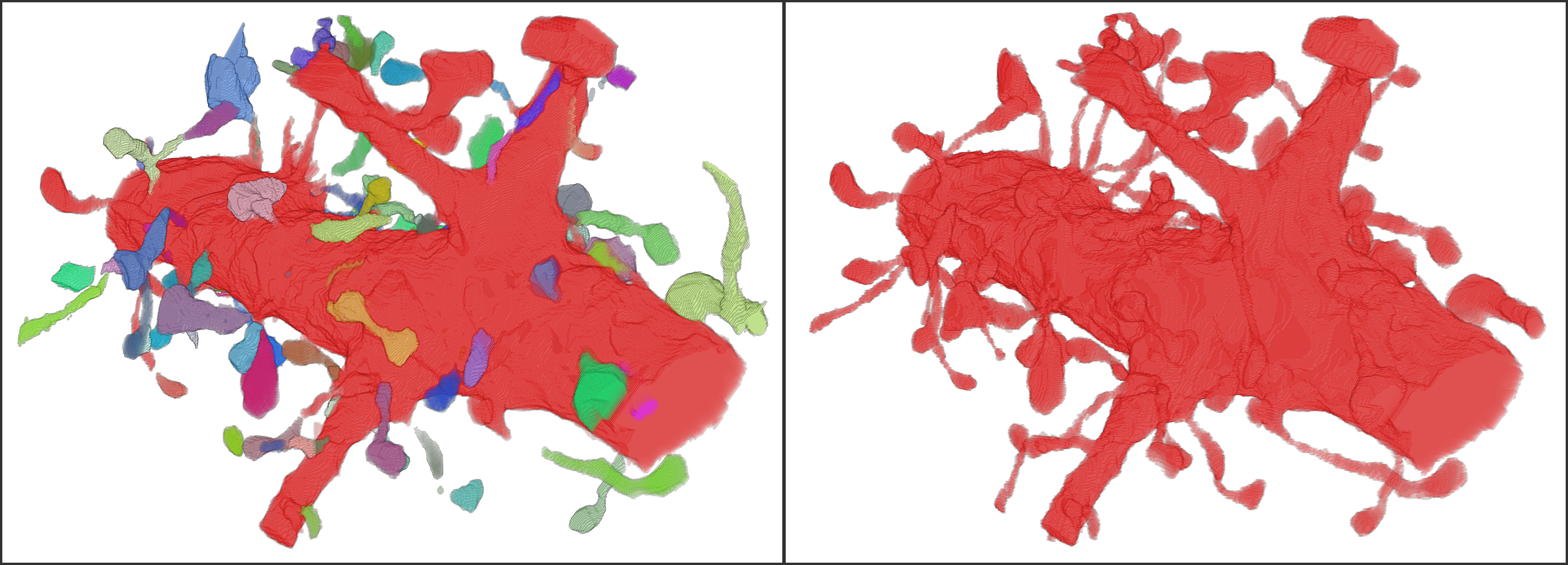}}
\caption{Automatic segmentation results (left) an manual annotations (right) for a 53.4{\textmu}m\textsuperscript{3} section of dendrite (13.1{\textmu}m in length).
131 merge operations (non-red colors) and 23 split operations (not shown) would be required to
proofread all errors larger than 0.0054 {\textmu}m\textsuperscript{3}.}\label{fig:splits}
\end{figure*}

\subsection*{Metrics}
It has been difficult to find consensus on the correct metric to use to measure segmentation accuracy in Connectomics. Easy to understand metrics such as ``error free path length'', ``edit distance'' and ``number of split and merge errors per cubic micron'' are favored by biologists because of the intuitive nature of these metrics. A short description of these metrics provides a good idea of what the resulting numbers mean and how close the segmentation comes to a correct result. Unfortunately these metrics are not robustly defined and require arbitrary decisions to be made about what constitutes an error.

For example, small disagreements between segmentations do not change the overall structure of the 3D reconstruction. Two annotations of the same volume made by different experts (or made by the same expert at different times) will contain many differences in exact boundary locations, even when the overall structure of the reconstructed objects is the same. Therefore, metrics such as ``errors per cubic micron" typically only count errors above a chosen minimum error volume. Similarly, the ``error free path length" metric will typically ignore small merge or split errors lasting for a chosen minimum path length or number of annotation nodes. Ignoring these small errors is a reasonable approach to take and necessary when using this type of metric, however it makes it difficult to robustly define the metric and prevent exploitation of tuning parameters in the metric.

To avoid uncertainty introduced by arbitrary decisions, the computer vision community favors metrics that count all errors in a contingency table and assign different weights to each error depending on the size of the error, and complexity of the ground truth segmentation. Rand index, Adjusted Rand index and variation of information (VI) are examples of this type of metric. Unfortunately these metrics result in a number which is difficult to interpret intuitively and does not provide an obvious link to how close the segmentation is to the ground truth or how much time it would take to correct the segmentation manually or with a semi-automated proof reading software.

Arganda-Carreras et al. addressed this problem by defining two evaluation metrics which on the one hand provide the rigorous definition needed for benchmarking, and on the other hand provide some intuition about the ratio of split and merge errors in the segmentation. The two metrics defined in their paper \citep{ArgandaCarreras2015ISBI} are based on the popular Rand and VI score and therefore named $V^\text{Rand}$ and $V^\text{Info}$, with the term Info referring to the information theoretic background of the definition of variation of information. For both metrics Arganda-Carreras et al. separate these metrics into split- and merge-focused sub-metrics ($V^\text{Rand}_\text{split}$, $V^\text{Rand}_\text{merge}$ and $V^\text{Info}_\text{split}$, $V^\text{Info}_\text{merge}$) which have a higher score if the segmentation contains fewer split (resp. merge) errors. As a summary metric they suggest the F-score, the harmonic mean between the split and the merge scores. Both metrics are normalized to a range between 0 and 1 with a higher value indicating a better segmentation. The evaluation done by Arganda-Carreras et al. shows that the ranking obtained with both metrics is not necessarily the same. While both metrics are normalized, they show different sensitivity to region sizes. The $V^\text{Rand}$ emphasizes correct segmentation of large structures, while $V^\text{Info}$ penalizes erroneous segmentation of smaller structures. 

Another interesting point is the influence of the background label. If pixels labeled as background either in the ground truth or the automatic segmentation are excluded from the evaluation, automated segmentations with wide background margins tend to produce more favorable scores. To avoid tuning of this unwanted behavior, it is now standard practice to ignore the background labels from the ground truth segmentation, and grow all regions from the automated segmentation until no pixels are labeled background anymore. We follow this practice and report scores for $V^\text{Rand}$ and $V^\text{Info}$ for all data sets as obtained from the RhoanaNet Pipeline segmentation. Note that extra-cellular space is considered background for the ECS dataset.

Another challenge for the Connectomics community is the difficulty in comparing techniques across different species, sample preparation techniques, imaging modalities, and ground-truth annotation methods. It is impossible to meaningfully compare results using any metric if the source data is not the same, and given the complexity of the image processing pipelines it is difficult to identify where one method performs better than another. Fortunately, open datasets and challenges such as the ISBI 2012 and 2013 neuron segmentation challenges, which used the S1 data set, provide a starting point for direct comparisons between segmentation methods. The RhoanaNet Pipeline goes one step further, enabling direct comparison of whole pipelines as well as methods addressing specific parts of the pipeline.

\section{Results}
\begin{figure}[!tpb]
\centerline{\includegraphics[width=\linewidth]{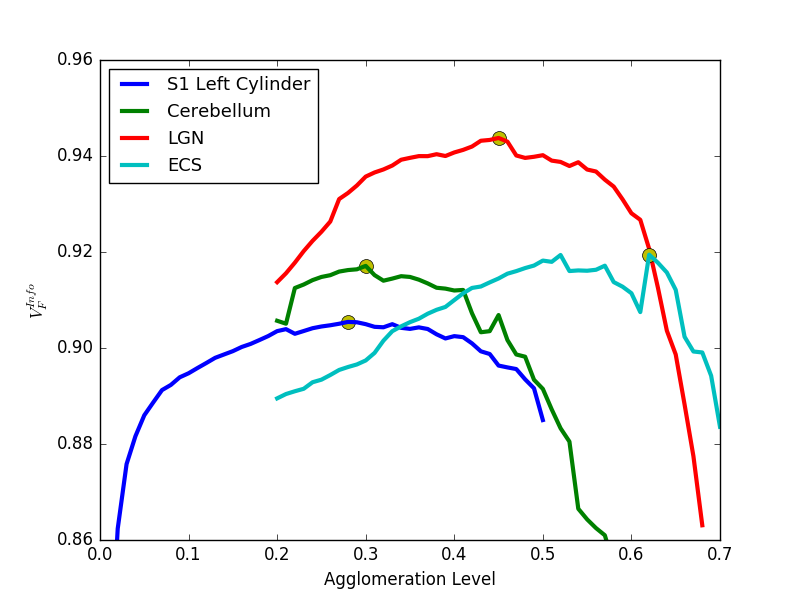}}
\caption{Segmentation metric $V^\text{Info}_\text{F-score}$ shown for a grid search over agglomeration levels. Maximal $V^\text{Info}_\text{F-score}$, results are highlighted with yellow dots with full results shown in Table \ref{Tab:0}.}\label{fig:f1plot}
\end{figure}

Here we demonstrate the RhoanaNet image processing pipeline on the four EM datasets discussed above. Deep neural networks were trained for each dataset individually, using the same network structure and hyper parameters each time. Agglomeration random forest training was performed on sub-volumes from the training data for each dataset individually, and cross-validation was used to choose the best random forest. A range of segmentation agglomeration levels were used to measure $V^\text{Info}_\text{F-score}$ on test data as shown in Figure \ref{fig:f1plot}. Full $V^\text{Rand}$ and $V^\text{Info}$ results for segmentations maximizing $V^\text{Info}_\text{F-score}$ are summarized in Table \ref{Tab:0}.

\subsection*{S1 Cortex}
This dataset contained very little noise, and Gala was able to connect most of the branching structures common in this data, including some spine necks, see Figure \ref{fig:splits}. A curiosity of this data set is that the training volume AC3 contains very little myelin. We therefore first trained on the whole training data (AC3 and AC4), and then fine tuned the network by restricting the training set to only images from AC4. We found that this training method lead to a slight increase in false positive detections on mitochondria, but a a good increase in true positive detection of myelinated cell boundaries. The decision to fine tune the network and all parameter tuning was performed on a validation set consisting of 10 images from AC3 and 10 images from AC4. No parameter tuning or additional training was performed based on the results on the test volume. 

\subsection*{Cerebellum}
The training set for this data is very small, but proved to be sufficient to train the U-Net to a satisfactory level. We follow the approach described by Ronneberger et al. \citep{Ronneberger2015} and use random rotations, flips, and non-linear deformations for data augmentation. 

\subsection*{LGN}
Despite lower contrast and the presence of some artifacts, U-Net training provided good boundary predictions for this dataset and the best agglomeration achieved the highest $V^\text{Info}_\text{F-score}$ of all datasets at 0.9438.

\subsection*{ECS Cortex}
This data set contains annotations from two different neurobiologists. Unfortunately the smaller set of annotations is restricted to a cropped volume of $632 \times 560 \times 40$ pixels. This size is smaller than the default configuration of the U-Net, which takes input patches of size $572 \times 527$ pixels. We therefore slightly reduced the size of the input patch for the U-Net to $540 \times 540$ pixels for this data set. This input size is still significantly larger than the context evaluated per pixel, thus it is unlikely to influence the accuracy of the network, but it slightly reduces the throughput performance during predictions.

\begin{table}[!t]
\begin{tabular}{lll>{\bfseries}lll>{\bfseries}l}
\toprule
~ & $V^\text{Rand}_\text{split}$ & $V^\text{Rand}_\text{merge}$ & \normalfont{$V^\text{Rand}_\text{F-score}$} & $V^\text{Info}_\text{split}$ & $V^\text{Info}_\text{merge}$ & \normalfont{$V^\text{Info}_\text{F-score}$}\\\midrule
S1 Cortex & 0.7850 & 0.9216 & 0.8478 & 0.9276 & 0.8843 & 0.9054 \\
Cerebellum & 0.9583 & 0.9731 & 0.9656 & 0.9253 & 0.9049 & 0.9150 \\
LGN & 0.9162 & 0.6705 & 0.7744 & 0.9590 & 0.9290 & 0.9438 \\
ECS Cortex & 0.9589 & 0.6170 & 0.7509 & 0.9718 & 0.8702 & 0.9182 \\
\bottomrule
\end{tabular}
\caption{Segmentation Results: Full segmentation metric results for each dataset.  F-scores for Rand and Information Theoretic metrics are in bold.}
\label{Tab:0}
\end{table}

\section{Discussion}
We presented the Rhoana pipeline for dense neuron annotation, updated to use deep learning U-Nets and region agglomeration using Gala. 3D segmentation results were qualitatively improved by the enhancements, with $V^\text{Info}_\text{F-score}$ scores ranging between 0.9 and 0.95 for the four datasets, and the pipeline can process very large volumes automatically.

High throughput of EM image data and quality of segmentation results will be very important for the future of Connectomics. Further improvements in segmentation quality and throughput are necessary and informed strategies for proofreading such as \citep{Plaza2012Minimizing,Haehn2014Design,Plaza2014Focused} will be required to minimize human effort. This open source pipeline provides an improvement in both quality and throughput and the open benchmark datasets provide an open and reproducible reference example on which future improvements can be built.

\section*{Acknowledgements}
Thanks to Linda Xu, students from Masconomet Regional High School and all our summer interns for their careful and diligent annotation work.

Also, thanks to Neal Donnelly, Princeton University, for contributions to the Gala project.

\paragraph{Funding:} We gratefully acknowledged support from NSF grants IIS-1447344 and IIS-1607800 and the Intelligence Advanced Research Projects Activity (IARPA) via Department of Interior/Interior Business Center (DoI/IBC) contract number D16PC00002, NIH/NINDS (1DP2OD006514-01, TR01 1R01NS076467-01, and 1U01NS090449-01), Conte (1P50MH094271-01), MURI Army Research Office (contract no. W911NF1210594 and IIS-1447786), CRCNS (1R01EB016411), NSF (OIA-1125087 and IIS-1110955), the Human Frontier Science Program (RGP0051/2014), Ruth L. Kirschstein Predoctoral Training Grant 5F31NS089223-02, and nVidia.

\bibliographystyle{apalike}
\bibliography{cmor}
\end{document}